# Analytical method for parameterizing the random profile components of nanosurfaces imaged by atomic force microscopy

Utkur Mirsaidov,*[a,b] Serge F. Timashev,[c,d] Yuriy S. Polyakov,[b] Pavel I. Misurkin,[e] Ibrahim Musaev[b,f] and Sergey V. Polyakov[b]

The functional properties of many technological surfaces in biotechnology, electronics, and mechanical engineering depend to a large degree on the individual features of their nanoscale surface texture, which in turn are a function of the surface manufacturing process. Among these features, the surface irregularities and self-similarity structures at different spatial scales, especially in the range of 1 to 100 nm, are of high importance because they greatly affect the surface interaction forces acting at a nanoscale distance. An analytical method for parameterizing the surface irregularities and their correlations in nanosurfaces imaged by atomic force microscopy (AFM) is proposed. In this method, flicker noise spectroscopy - a statistical physics approach - is used to develop six nanometrological parameters characterizing the high-frequency contributions of jump- and spike-like irregularities into the surface texture. These contributions reflect the stochastic processes of anomalous diffusion and inertial effects, respectively, in the process of surface manufacturing. The AFM images of the texture of corrosion-resistant magnetite coatings formed on low-carbon steel in hot nitrate solutions with coating growth promoters at different temperatures are analyzed. It is shown that the parameters characterizing surface spikiness are able to quantify the effect of process temperature on the corrosion resistance of the coatings. It is suggested that these parameters can be used for predicting and characterizing the corrosion-resistant properties of magnetite coatings.

## 1. Introduction

The functional properties of coatings for biochip substrates,[1] semiconductor thin films,[2] ultrafine-grained titanium surfaces for bacterial attachment,[3] silicon-wafer surfaces for neural cell attachment,[4] magnetite coatings on low-carbon steel,[5] and many other technological surfaces depend to a large degree on their nanoscale surface texture, which in turn is a function of their manufacturing process. This implies that the nanometrological parameters obtained from texture measurement and analysis of such surfaces should adequately relate their functional properties and manufacturing conditions, identifying and characterizing the important features of the surface topography.[6,7] Among these features, the surface irregularities and self-similarity structures at different spatial scales, especially in the range of 1 to 100 nm, are of high importance because they greatly affect the surface interaction forces acting at a nanoscale distance, which in turn control the values of physicochemical parameters such as binding energy, adhesion, resistance to abrasion, and the like. Consequently, the texture of these surfaces should be measured with atomic resolution.

The most versatile tool to make such measurements is atomic force microscopy (AFM), which can provide three-dimensional high-resolution images of different objects in air, liquid, or vacuum environments:[6,8,9] proteins,[10] DNA molecules,[11] nanoholes (biosensing),[12] coatings for biochip substrates,[1] cell surfaces,[13] and semiconductor thin films.[2] AFM measurements provide series of recorded digitized values of surface profile heights $h(x; y)$,[6,8,9] which are produced with a sensitive probe fixed at different values of coordinate $y$ and scanning along the coordinate $x$, normal to $y$, in the range $0 \leq x \leq L$ for each line scan ($L$, characteristic sample dimension for digitized image). In this case, all the information to be extracted refers to the range of spatial frequencies $1/L < f < 0.5\ f_d$ ($f_d$, spatial discretization frequency, with typical values of $L \sim$ 1-10 μm and $f_d = 1/\Delta l \sim$ 1 nm$^{-1}$, where $\Delta l$ is the elementary probe step size).

There are two major approaches to surface texture analysis in nanometrology: profile (2D) and areal (3D).[6,7] Profile surface characterization studies the variations of surface heights along a selected coordinate of the surface. Areal surface texture analysis examines the height variations for surface areas and studies the topographic features of surface texture. The conventional scheme for both profile and areal surface texture analyses, which is used for small-scale surfaces in the range from nano- to millimeter, typically includes three main steps: fitting, filtering, and parameterization.[6,7] At the fitting (preprocessing) stage, measurement setup and part-induced errors, such as background slopes due to the tilt of the sample on the sample holder in AFM or the curvature of scanned cylindrical surfaces, are rectified using linear algebra, matrix, or non-linear optimization procedures. At the filtering stage, the measured surface profile or area is partitioned into different wavelength bandwidths or scale-limited surfaces using, as a rule, a variant of the Gaussian filter or more advanced spline,

Gaussian regression, or wavelet-based filters. At the parameterization stage, integral (statistical) parameters are calculated either for each bandwidth (roughness or waviness) in the case of profile analysis or for scale-limited surfaces in the case of areal analysis. Some of the standard roughness parameters (ISO 4287:1997) are the average roughness *Ra*, root mean square roughness *Rq*, skewness *Rsk*, kurtosis *Rku*, and profile maximum height *Rz* and their areal generalizations (ISO 25178): arithmetic mean height of the surface *Sa*, root mean square height of the surface *Sq*, skewness of height distribution *Ssk*, kurtosis of height distribution *Sku*, and surface maximum height *Sz*. Areal parameterization also includes feature characterization, pattern analysis of surface texture by identifying and parameterizing the topographical features of specific types.

The general filtering and parameterization procedures outlined above were developed for the analysis of an arbitrary small-scale surface, with heights typically ranging from nanometers up to a millimeter. At the same time, nanoscale surfaces have some distinctive features separating them out from micro- or millimeter-scale surfaces. At the nanoscale, the structure of molecules and the interactions between them start controlling surface texture, resulting in sharp changes in roughness. For example, it was shown that efficient dissociative adsorption of diatomic molecule $H_2$, a central step in many industrial catalytic processes, on a palladium catalyst surface requires three or more adjacent and empty adsorption sites (vacancies), which implies a preparatory surface reconfiguration at the scales of up to few nanometers.[14] Adsorption of hydrogen atoms on a clean metal or semiconductor surface is known to result in significant changes of self-diffusion rates (thin-film growth and sintering processes), which leads to major structural reconfiguration of nanoscale surface fragments; for example, an activated Pt-H complex enhanced the diffusivity by a factor of 500 at room temperature, relative to other Pt adatoms (without hydrogen).[15] The studies of singlet oxygen photogeneration (photodynamic therapy) upon photoexcitation of deposited porphyrin layers demonstrated that the structure of deposited layers on scales of up to dozens of nanometers, which depends on the chemical structure of porphyrin molecules, is a major factor in the process of photogeneration.[16]

The above implies that surface irregularities, which are related to intermolecular interactions and molecular structures, may account for some functional properties specific to nanosurfaces. The effect of surface irregularities on the functional properties depends on the structural organization of individual molecular complexes into surface fragments and the correlations between molecular complexes/surface fragments. The information about surface irregularities and their correlations at scales up to 100 nm is most relevant for the problems of catalysis, electrochemistry and corrosion, adhesion, and tribology. Of special significance for surface chemical activity are spike-like (needle-shaped) nanoscale irregularities that are associated with high local values of electric-field intensity and mechanical stress.[17] It should be noted that the functional role of correlations in recorded surface heights, which are not captured with standard statistical parameters, such as *Ra*, *Rq*, *Rsk*, *Rku*, *Rz*, *Sa*, *Sq*, *Ssk*, *Sku*, and *Sz*, and cannot be filtered using band pass filters (almost all scales of a specific 3D image should be considered), was already explored in surface texture analysis studies based on fractal and multifractal methods.[6,18,19]

To study the effects of surface irregularities and their correlations on the functional properties, which are not discussed in ISO 4287:1997 and ISO 25178, additional methods and parameters specific to nanosurfaces need to be introduced into nanoscale surface texture analysis. An analytical method of surface texture analysis, which deals with surface areas and combined profiles accounting for topographical variations in surface segments, is proposed in this paper on the basis of flicker noise spectroscopy (FNS), a statistical physics framework for the analysis of time and space series.[19-23] To analyze combined surface profiles with randomly varying components, the method extracts information from the series of surface height irregularities by separating out the rapid height changes of different types, called "jumps" and "spikes", which are present at different surface scales, from the background profile with slow height changes at scales of one to two orders smaller than the linear size of the image under study. In this case, the introduced stochastic information parameters characterize the "measure" of these irregularities and the correlations found in the series of recorded digitized heights $h(x; y)$. The values of FNS parameters for an AFM image or its fragments are determined on the basis of partitioning the array of measured surface profile heights corresponding to a 3D surface into stripes and averaging over multiple line scans.

The section 2 of this paper briefly presents FNS principles and an algorithm implementing the proposed method. The section 3 illustrates the texture analysis of nanosurfaces by the proposed method.

## 2. Method

### 2.1. FNS principles

Here, we will deal only with the basic FNS relations needed to understand the principles of constructing the method to be proposed. FNS is described in more detail elsewhere.[20-24] In FNS, it is assumed that the correlation between the new profile height values $h(x + \Delta)$, $\Delta > 0$, and the old ones can be used to extract the information contained in the heights $h(x)$ of each surface profile. In this case, the local values of $h(x)$ are defined as the dynamic variables of the system under study and the surface parameters are related to the autocorrelation function, one of the basic functions in statistical physics, in analyzing every surface profile:

$$\psi(\Delta) = \langle h(x) h(x+\Delta) \rangle_{L-\Delta} \qquad (1)$$

where $\Delta$ is the spatial shift and the angular brackets stand for the averaging over the interval $L - \Delta$:

$$\langle (...) \rangle_{L-\Delta} = \frac{1}{L-\Delta} \int_0^{L-\Delta} (...) \, dx \qquad (2)$$

The function $\psi(\Delta)$ characterizes the relationship between the profile heights $h(x)$ at large and small argument values. The averaging over the interval $L$ implies that all the characteristics that can be extracted by analyzing the function $\psi(\Delta)$ should be regarded as averaged over this interval. In this case, each surface profile $h(x)$ is represented as a basic low-frequency profile against the background of which random components are identified. These components can be represented as series of jump-like random values of variable $h(x)$, which account for the "diffusional" (in the general case, non-Fickian) dispersion of the basic profile,[21] and higher-frequency spike-like random values, which are caused by the inertial effects taking place in the formation of the profile.

The information contained in $\psi(\Delta)$ is taken out by analyzing two "projections" of this function: the "incomplete" cosine transform $S(f_x)$ of the autocorrelation function (power spectrum estimate),

$$S(f_x) = \int_0^L \langle h(x)h(x+x_1)\rangle_{L-x_1} \cos(2\pi f_x x_1) dx_1 \quad (3)$$

(here, we assume $\langle h(x)\rangle = 0$, where $f_x$ is the spatial frequency), and the difference moment of the second order $\Phi^{(2)}(\Delta)$ (Kolmogorov transient structural function),

$$\Phi^{(2)}(\Delta) = \langle [h(x) - h(x+\Delta)]^2 \rangle_{L-\Delta} \quad (4)$$

The random component of $\Phi^{(2)}(\Delta)$ is based solely on the jump-like random irregularities in the dynamic variable at each hierarchical level of the system whereas $S(f_x)$ is based not only on jump-like random irregularities but also on more rapidly varying spike-like random irregularities.[20-21]

The basic idea in parameterization is to use two three-parameter interpolation expressions for the random components. The first expression is used to determine the spectral contribution of the random components of surface profile heights $h(x)$ and exclude the contribution of the low-frequency component to the parameters related to jump- and spike-like random irregularities. The second interpolation, which deals with the random component in the structural function, is used to determine the parameters characterizing the series of jump-like random irregularities in the surface profile, which correspond to anomalous diffusion.

As many as 6 FNS parameters are introduced to characterize the randomness in the surface profile.[20,21] The first parameter is $\sigma$, the root mean square deviation of the value of the measured dynamic variable from the slowly varying basic surface profile, which is used as a surface-profile stepwiseness factor and based solely on jump-like random irregularities. The second parameter accounts for the "intensity" of jump- and spike-like random irregularities in the highest-frequency interval $f_x \sim 0.01\text{-}1$ nm$^{-1}$, where many functional properties of solid surfaces are identified. It can be represented by the value of $S_c(f_x)$ when $f_x \sim L_0^{-1}$, where $L_0$, the third FNS parameter, is the length of correlation for the high-frequency irregularities at this nano-scale. The parameter $S_c(L_0^{-1})$ will be called the texture spikiness factor. The other three parameters include the Hurst constant $H_1$, which characterizes the rate at which the dynamic variable loses the "memory" about its value in spatial intervals smaller than the correlation length $L_1$ (here, the fifth FNS parameter $L_1$ can be interpreted as the characteristic distance at which the dispersion of measured surface profile heights is formed), and the flicker noise parameter $n$, which characterizes the rate of loss of correlations in the series of high-frequency irregularities in spatial intervals $L_0$.

The stochastic parameters are determined by the following procedure. First, the original AFM image is partitioned into multiple stripes and the surface height profiles are averaged for each stripe to take into account topographical variations. Second, the fitting and discrete transform techniques are used to separate out random components from the combined profiles and calculate the values of the FNS parameters.

## 2.2. Partitioning of AFM image into multiple stripes

The original AFM image for a $L \times L$ sample is a two-dimensional array $h(x_i, y_j)$, $(i = 1, 2, .., N; j = 1, 2, .., N)$ of surface height values.

1. The array is partitioned over the axis $y$ (bottom-up) into $m$ stripes, parallel to the axis $x$, of equal size: $h_k(x_i, y_j)$, $(i = 1, 2, .., N; j = 1, 2, .., N/m; k = 1, 2, ..., m)$.
2. For each $k$-th stripe, we calculate a series of averaged values of heights along a selected axis (in this case we use the axis $x$):

$$h_k(x_i) = \frac{m}{N} \sum_{j=1}^{N/m} h_k(x_i, y_j) \quad (5)$$

3. Each of the above series of numbers $h_k(x_i)$ is interpreted as a signal, for which the FNS parameters $P_k = \{\sigma, L_0, L_1, H_1, n_0, S_c(L_0^{-1})\}_k$, $(k = 1, 2, ..., m)$ are calculated using the algorithm given in the subsection 2.3.
4. The resulting parameters $P$ for the surface are calculated by the formula:

$$P = \{\sigma, L_0, L_1, H_1, n_0, S_c(L_0^{-1})\} = \frac{1}{m}\sum_{k=1}^{m} P_k \quad (6)$$

where $P_k = \{\sigma, L_0, L_1, H_1, n_0, S_c(L_0^{-1})\}_k$.

## 2.3. Parameterization of combined roughness profile

Consider a spatial series $h(x_i)$ (subscript $k$ in $h_k(x_i)$ is dropped for simplicity). In this case, the parameterization procedure can be written as follows:

1. Calculate the arithmetic mean for the signal:

$$\mu_h = \frac{1}{N}\sum_{i=1}^{N} h(x_i) \quad (7)$$

2. Subtract the arithmetic mean from the series $h(x_i)$:

$$\bar{h}(x_i) = h(x_i) - \mu_h \quad (8)$$

3. Calculate the autocorrelation function for the series $\bar{h}(x_i)$:

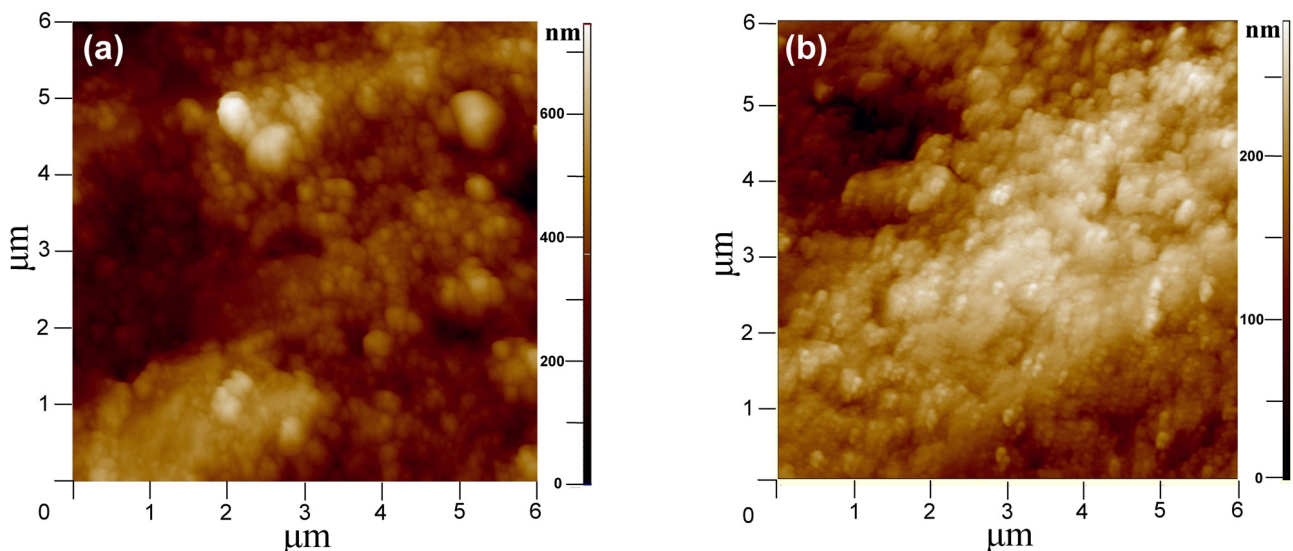

Fig. 1 AFM micro-images of magnetite coatings formed at (*a*) 70 and (*b*) 98°C.

$$\psi_d(p) = \frac{1}{N-p}\sum_{i=1}^{N-p} \bar{h}(x_i)\bar{h}(x_{i+p}) \quad \text{when} \quad p = 0..M \quad (9)$$

Let the autocorrelation interval $M$ be $N/4$ (higher values of $M$ will result in the loss of statistical information in estimating the autocorrelation function). To go from discrete form to the continuous one, one can use the following expression: $p = N\Delta/L$, where $\Delta$ is the probe step size. The index $d$ here and below is used to denote the discrete form of expressions.

4. Calculate the discrete cosine transform of the autocorrelation function:

$$S_d(q) = \psi_d(0) + \psi_d(M)(-1)^q + 2\sum_{p=1}^{M-1}\psi_d(p)\cos\left(\frac{\pi q p}{M}\right) \quad (10)$$

where $q = 0..M$. For $q = 1..M-1$, $S_d(q)$ should be multiplied by 2, which is the standard procedure for discrete Fourier transforms to take into account the spectral values in the second half of the frequency range. Here, relations $q = 2f_x f_d^{-1} M$ and $S_d(q) = S(f_x) \times f_d$ describe the equivalence between the discrete and continuous forms of power spectrum estimate.

5. Calculate $S_{cd}(0)$ as the average value of the power spectrum for the points 2 and 3 (point 1, which corresponds to the zero frequency, is not used in calculating $S_{cd}(0)$):

$$S_{cd}(0) = \frac{S_d(1)+S_d(2)}{2} \quad (11)$$

6. Interpolate $|S_d(q)|$ using the expression:

$$S_{cd}(q) = \frac{S_{cd}(0)}{1+(\pi\frac{q}{M}L_{0d})^n} \quad (12)$$

by the method of nonlinear least-square fitting to determine the values of parameters $n$ and $L_{0d}$. The fitting is done on the basis of a double logarithmic scale, dividing the entire series into a set of equal intervals. We used the trust-region algorithm for nonlinear square fitting,[25] which is built in MATLAB v.7 or higher.

7. Separate out the resonant component:

$$S_{rd}(q) = S_d(q) - S_{cd}(q) \quad \text{when} \quad q = 0..M \quad (13)$$

8. Calculate the autocorrelation function for the resonant component as the inverse discrete cosine transform of $S_{rd}(q)$. When $q = 1..M-1$, divide $S_{rd}(q)$ by 2 to take into account the spectral values in the second half of the frequency range. Then calculate the inverse cosine transform:

$$\psi_{rd}(p) = \frac{1}{2M}\left(S_{rd}(0) + S_{rd}(M)(-1)^p + 2\sum_{q=1}^{M-1}S_{rd}(q)\cos\left(\frac{\pi p q}{M}\right)\right) \quad (14)$$

9. Calculate the difference moment for the resonant component:

$$\Phi_{rd}^{(2)}(p) = 2[\psi_{rd}(0) - \psi_{rd}(p)] \quad \text{when} \quad p = 0..M \quad (15)$$

The continuous equivalent of $\Phi_{rd}^{(2)}(p)$ is $\Phi_{r}^{(2)}(\Delta)$.

10. Calculate the difference moment for the experimental series:

$$\Phi_d^{(2)}(p) = \frac{1}{N-p}\sum_{i=1}^{N-p}\left[\bar{h}(x_i) - \bar{h}(x_{i+p})\right]^2 \quad \text{when} \quad p = 0..M \quad (16)$$

11. Calculate the difference moment for the random component:

$$\Phi_{ecd}^{(2)}(p) = \Phi_d^{(2)}(p) - \Phi_{rd}^{(2)}(p) \quad \text{when} \quad p = 0..M \quad (17)$$

The continuous equivalent of $\Phi_{ecd}^{(2)}(p)$ is $\Phi_{ec}^{(2)}(\Delta)$.

12. Determine the parameters $\sigma$, $H_1$, $L_{1d}$ by fitting $\Phi_{ecd}^{(2)}(p)$ in Eq. (17) to the interpolation expression of the anomalous

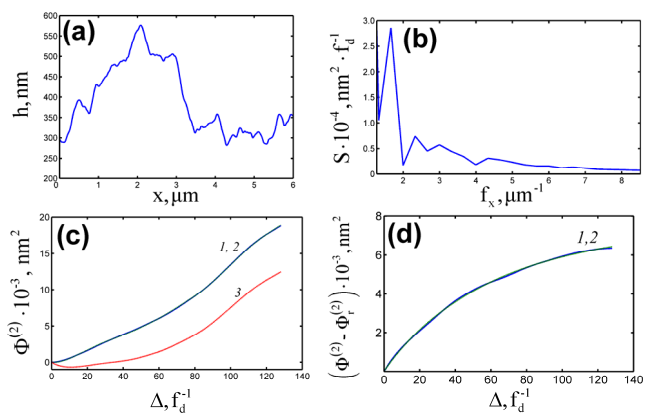

**Fig. 2** FNS texture analysis of the AFM image for the magnetite coating formed at 70°C: (*a*) averaged (over *y*) height $h(x)$ for stripe 2; (*b*) $S(f)$ at low spatial frequencies, Eq. (10); (*c*) structural functions $\Phi^{(2)}(\Delta)$: *1*, experimental, Eq. (16); *2*, calculated, Eq. (15) + Eq. (18); *3*, contribution of resonant frequencies, Eq. (15); (*d*) stochastic component $\Phi_{ec}^{(2)}(\Delta)$: *1*, experimental, Eq. (17); *2*, calculated, Eq. (18).

diffusion type:[20]

$$\Phi_{cd}^{(2)}(p) = 2\sigma^2 \times \left[1 - \Gamma^{-1}(H_1) \cdot \Gamma(H_1, p/L_{1d})\right]^2 \quad (18)$$

where $\Gamma(s,x) = \int_x^\infty \exp(-t) \cdot t^{s-1} dt$, $\Gamma(s) = \Gamma(s,0)$, using the same least-square fitting method as in step 6. The continuous equivalent of $\Phi_{cd}^{(2)}(p)$ is $\Phi_c^{(2)}(\Delta)$.

13. Calculate $S_{cd}(L_{0d}^{-1})$ by Eq. (12).

14. After the values of all six FNS parameters – $\sigma$, $L_{0d}$, $L_{1d}$, $H_1$, $n$, $S_{cd}(L_{0d}^{-1})$ - are determined, calculate the dimensional values for $L_{0d}$, $L_{1d}$, $S_{cd}(L_{0d}^{-1})$: $L_0 = L_{0d} \times \Delta l$, $L_1 = L_{1d} \times \Delta l$, $S_c(L_0^{-1}) = S_{cd}(L_{0d}^{-1}) \times \Delta l$.

15. Calculate the relative error $\varepsilon_\Phi$ in the interpolation of difference moment $\Phi_d^{(2)}(p)$:

$$\varepsilon_\Phi = \frac{\sum_{p=1}^{M}\left|\Phi_d^{(2)}(p) - \Phi_{rd}^{(2)}(p) - \Phi_{cd}^{(2)}(p)\right|}{\sum_{p=1}^{M}\Phi_d^{(2)}(p)} \times 100\% \quad (19)$$

Here, the error is determined as the ratio of the difference of areas between the experimental structural function and the total interpolation function to the area of the experimental structural function. The areas are calculated by numerical integration using the rectangle method because the original series have a rather large number of points. The parameterization is successful if $\varepsilon_\Phi \leq 10\%$.[21]

## 3. Analysis of AFM images for magnetite coatings on low-carbon steel

### 3.1. Experimental

We illustrate the proposed method for parameterizing the random components of nanosurfaces by analyzing the texture of magnetite coatings (MC) on low-carbon steel, which are formed in hot nitrate solutions using a previously reported method.[5] In this method, a URMP-5 magnetite coating growth promoter containing nitrates of metals with small cation radii is added to a solution of ammonium nitrate, which provides high protective properties of MCs at oxidation temperatures less than 100°C. It is shown below how the values of FNS parameters account for the changes in the complex MC texture formed during 40 min in a 25 g/L $NH_4NO_3$ solution mixed with a 0.1 g/L URMP-5 promoter at 70 and 98°C. An INTEGRA-TERMA scanning probe microscope, manufactured by NT-MDT (Zelenograd), was used to obtain AFM images of the MC textures. Images of the texture fragments of the formed coatings were made in the tapping mode using an NSG-11 silicon cantilever of type B with a resonance frequency of 150 kHz and a tip curvature radius of 10 nm. The number of line scans was 512 with 512 points in each line scan, producing a total 512×512 array of measured points. The AFM images of oxidized 6×6 μm samples are illustrated in Fig. 1.

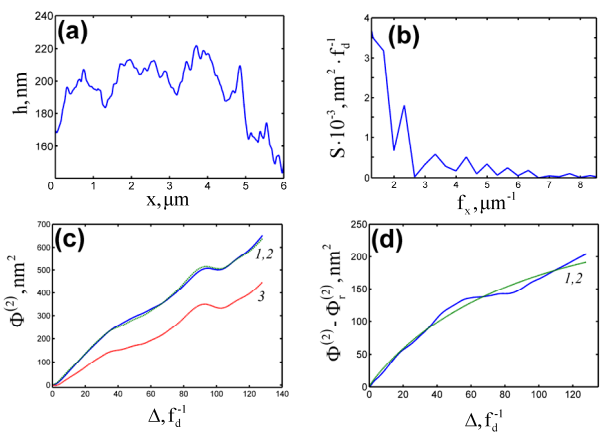

**Fig. 3** FNS texture analysis of the AFM image for the magnetite coating formed at 98°C (stripe 3): nomenclature as in Fig. 2.

### 3.2. Analysis

As most of instrumentation errors and environmental noise are usually accounted for before parameterizing AFM images (at the preprocessing stage)[6], the effects of tip topography, tip artifacts, tip surface contamination, AFM setup, and environmental conditions during imaging (external noise, humidity, temperature) are beyond the scope of this paper.

According to the partitioning algorithm, each set of 512 line scans forming the image was partitioned normal to line scans into 8 stripes of 64 rows of height points each (1 to 64, 65 to 128, etc.). The number of stripes depends on the problem under study. The general rule is that the stripe width should not be less than 10 line scans; otherwise, the topographical variations would not be captured. On the other hand, the upper bound should be close to the correlation length $L_0$. Calculations at different numbers of stripes should generally be performed to see if FNS parameters significantly change for different numbers of stripes. If that is the case, then the parameter values at different numbers of stripes should be used in the analysis of the relation between FNS parameters and functional properties. For each stripe, we determined the combined surface profiles $h_k(x)$ ($k = 1, 2, 3,$

**Table 1** FNS parameters for the magnetite coating formed at 70°C

| Stripe | $Ra$, nm | $\delta$, nm | $\sigma$, nm | $H_1$ | $L_1$, µm | $S_c(L_0^{-1})$, nm²·µm | $L_0$, µm | $n$ | $\varepsilon_\Phi$ |
|---|---|---|---|---|---|---|---|---|---|
| 1 | 394 | 97 | 47 | 0.55 | 0.23 | 98 | 0.39 | 1.89 | 1.78% |
| 2 | 392 | 102 | 61 | 0.47 | 0.98 | 206 | 0.64 | 1.84 | 0.41% |
| 3 | 299 | 88 | 36 | 0.91 | 0.39 | 43 | 0.48 | 2.49 | 0.61% |
| 4 | 316 | 87 | 27 | 1.16 | 0.07 | 99 | 0.52 | 1.84 | 8.54% |
| 5 | 313 | 96 | 29 | 1.97 | 0.05 | 112 | 0.54 | 1.92 | 6.98% |
| 6 | 337 | 91 | 54 | 0.83 | 0.54 | 56 | 0.39 | 2.44 | 0.27% |
| 7 | 412 | 107 | 46 | 0.79 | 0.41 | 146 | 0.74 | 2.10 | 0.63% |
| 8 | 347 | 104 | 22 | 0.77 | 0.08 | 128 | 0.53 | 1.84 | 2.71% |
| *Average* | 351 | 96 | 40 | 0.93 | 0.34 | 111 | 0.53 | 2.05 | 2.74% |

**Table 2** FNS parameters for the magnetite coating formed at 98°C

| Stripe | $Ra$, nm | $\delta$, nm | $\sigma$, nm | $H_1$ | $L_1$, µm | $S_c(L_0^{-1})$, nm²·µm | $L_0$, µm | $n$ | $\varepsilon_\Phi$ |
|---|---|---|---|---|---|---|---|---|---|
| 1 | 142 | 25.1 | 5.3 | 1.00 | 0.05 | 1.77 | 0.49 | 3.01 | 3.22% |
| 2 | 169 | 26.5 | 9.9 | 0.45 | 0.81 | 1.78 | 0.61 | 1.90 | 2.85% |
| 3 | 194 | 24.3 | 11.2 | 0.44 | 1.44 | 2.04 | 0.39 | 1.88 | 1.71% |
| 4 | 213 | 30.0 | 14.4 | 0.84 | 1.15 | 2.07 | 0.44 | 2.68 | 0.58% |
| 5 | 192 | 44.9 | 17.1 | 0.46 | 0.25 | 1.85 | 0.50 | 1.92 | 0.92% |
| 6 | 180 | 59.5 | 14.7 | 0.96 | 0.09 | 1.76 | 0.66 | 2.91 | 4.49% |
| 7 | 158 | 62.8 | 27.9 | 0.90 | 0.15 | 1.89 | 0.45 | 2.79 | 6.76% |
| 8 | 165 | 34.4 | 4.4 | 1.16 | 0.04 | 2.32 | 0.42 | 3.32 | 5.48% |
| *Average* | 177 | 38.4 | 13.1 | 0.78 | 0.50 | 1.93 | 0.50 | 2.55 | 3.25% |

…, 8), where the subscript $k$ is the stripe number measured from the bottom of each image given in Fig. 1. In this case, we considered stripes parallel to the $x$ axis. It is also possible to analyze stripes parallel to the $y$ axis if the line scans are equally spaced from each other. The combined surface profiles calculated in the interval $0 \leq x \leq L$ for stripes 2 (70°C) and 3 (98°C) are plotted in Figs. 2a and 3a, respectively. The values of FNS parameters calculated by the above parameterization procedure are listed in Tables 1 and 2.

Figures 2b and 3b illustrate the spectral functions $S(f_x)$ in the low-frequency section of the spectrum, which were used to calculate the resonance contributions $\Phi_r^{(2)}(\Delta)$ into the structural function. The structural functions $\Phi^{(2)}(\Delta)$ calculated by Eq. (16) for the given profiles and those plotted as the sum of the resonant components $\Phi_r^{(2)}(\Delta)$ calculated by Eq. (15) and the random components $\Phi_c^{(2)}(\Delta)$ calculated by Eq. (18) are shown in Figs. 2c, 3c. In these figures, the value of $\Delta$ is given in relative units (r.u.): 1 r.u. = 6 µm /512 ≈ 0.0117 µm.

The figures demonstrate a good agreement between the experimental and calculated (by nonlinear square fitting)[25] values of $\Phi^{(2)}(\Delta)$, which is achieved for every studied roughness profile $h_k(x)$ at certain values of stochastic parameters $\sigma$, $H_1$, and $L_1$. The fact that the random components $\Phi_{ec}^{(2)}(\Delta)$ are almost perfectly interpolated by expression (18), which follows from Figs. 2d and 3d, implies that the formation of jump-like irregularities is controlled by the stochastic process of anomalous diffusion.

In addition to the calculated values of FNS parameters, Tables 1 and 2 present two additional surface stripe characteristics of the images: (1) the values of $Ra$ for every stripe and (2) the corresponding standard deviations - $\delta$. Among the six FNS parameters, the three parameters characterizing spike-like random irregularities, $S_c(L_0^{-1})$, $L_0$, and $n_0$, are of special interest for the analysis of MC AFM images. It was shown that the corrosion activity of MCs is related to the integral characteristics of spike-like irregularities because mechanical stresses are concentrated at the points where the surface profile rapidly changes and this can be accompanied by the emergence of local electrical fields of high strength, which can activate corrosion processes.[5]

The comparison between the values of FNS parameters listed in Tables 1 and 2 shows that the MC formed at 98°C is characterized by $S_c(L_0^{-1})$ values much less than those for the MC formed at 70°C. This ratio of ≈58 is much larger than the ratio between the squares of average values $Ra$ (≈4), which are proportional to the absolute values of corresponding power spectrum estimates. At the same time, we can see that the corresponding ratios for stepwiseness factor $\sigma$ and $Ra$ are only slightly different. This implies that the relative intensity of spike-like irregularities drops by one order of magnitude for the MC formed at 98°C while the relative intensity of jump-like irregularities stays practically the same. The validity of expressions $n \approx 2H_1 + 1$ and $L_1 \approx L_0$ at 98°C (Table 2), which corresponds to the case when the diffusion component (jump-like irregularities) dominates over the inertial component (spike-like irregularities),[20-22] implies that the effect of spike-like irregularities related to corrosion susceptibility becomes insignificant. Therefore, the six FNS parameters for the MC surface at 98°C get reduced to three jump-related parameters $\sigma$, $L_1$, and $H_1$. The value of correlation length $L_1$ increases by 47% and Hurst constant $H_1$ declines by 16%. Hence, the correlations in jump-like (diffusion) random irregularities fade away slower at 98°C. The above analysis demonstrates that the intensity of spike-like irregularities dramatically decreases when the temperature is increased from 70 to 98°C, suggesting that the MC formed at 98°C should show up higher protective properties, which is actually confirmed by their corrosion tests.[5]

It should also be noted that as the texture of both MC samples was rather heterogeneous, the scatter of the values of parameter $S_c(L_0^{-1})$ for different areas of the MC surface can be regarded as a quality factor for the corrosion-resistant coating. In terms of the scatter of these parameters, the MC formed at 70°C demonstrates higher texture heterogeneity as compared to the MC formed at 98°C.

The purpose of the above analysis is to illustrate the proposed mathematical method rather than perform a comprehensive study of the relation between FNS parameters and MC corrosion resistance properties. In the latter case, when an accurate table mapping the values of FNS parameters to MC corrosion resistance properties for a specific experimental setup is to be prepared, one would need to analyze the AFM images of control surfaces and consider different instrumentation errors (tip topography, tip contamination, tip artifacts, and others) and environmental conditions during the experiment (external noise, humidity, temperature).

## 4. Conclusions

The proposed method of surface texture analysis, which deals with surface areas and combined AFM profiles, made it possible to separate out and parameterize the high-frequency contributions of jump- and spike-like irregularities into the nanosurface texture of corrosion-resistant coatings, which are related to the stochastic processes of anomalous diffusion and inertial effects, respectively, in the process of coating manufacturing. The analysis based on the six FNS parameters developed in this paper showed that the parameters characterizing surface spikiness may account for the corrosion resistance of magnetite coatings on low-carbon steel.

The proposed FNS parameterization of technological surfaces at nanoscales may be used to analyze the texture quality of other technological materials with random surface irregularities (electronic conductors, frictional and adhesional contacts, corrosion-resistant and catalytic coatings, thin semiconductor films, coatings for biochip substrates, etc.) the functional properties of which depend to a large degree on the individual features of their nanoscale surface texture. The methodology of FNS texture analysis can be generalized for separating out and parameterizing the random components and their correlations in the physicochemical characteristics (magnetic, electrical, elastic, etc.) measured by scanning probe microscopy, which characterize the functional properties of various technological materials. The method can be adopted for parameterizing the texture images made by optical methods, for example, to analyze the sections of a biological tissue in order to diagnose its condition. It can be of much help in analyzing the images of synthetic biological tissues and biofilms synthesized with optical tweezers and a microfluidic.[26,27]

## Acknowledgements

The authors would like to thank Yu.I. Kuznetsov, A.B. Solovyeva, and D.B. Vershok for the preparation of magnetite coating samples and discussion of the results of FNS analysis of their texture.

## Notes and references

[a] RCE in Mechanobiology, National University of Singapore, 5A Engineering Drive 1, Singapore, 117411; E-mail: mirsaidov@gmail.com; Fax: +65 68726123; Tel.: +65 65161689
[b] USPolyResearch, 906 Spruce St., Ashland, PA, 17921, USA; E-mail: ypolyakov@uspolyresearch.com; Tel.: +1 347 6737747
[c] Karpov Institute of Physical Chemistry, Ul. Vorontsovo pole 10, Moscow 103064, Russia
[d] Institute of Laser and Information Technologies, Russian Academy of Sciences, Pionerskaya str. 2, Troitsk, Moscow Region, 142190, Russia
[e] Semenov Institute of Chemical Physics, Russian Academy of Sciences, Ul. Kosygina 4, Moscow 19991, Russia
[f] Yale School of Management, Yale University, Box 208200, New Haven, CT, 06520, USA